\documentstyle[12pt]{article}
 \textwidth
 16.4cm
 \oddsidemargin
 2.5cm
 \advance\oddsidemargin
 by
 -1in
 \evensidemargin
 0.0cm
 \advance\evensidemargin
 by
 -1in
 \marginparwidth
 1.9cm
 \marginparsep
 0.4cm
 \marginparpush
 0.4cm
 \topmargin
 -1.5cm
 \advance\topmargin
 by
 -0.0in
 \textheight
 23cm
 \makeindex

 \newcommand\la{\langle}
 \newcommand\ra{\rangle}
 \newcommand\noi{\noindent}
 \newcommand\beq{\begin{equation}}
 \newcommand\eeq{\end{equation}}
 \newcommand\beqn{\begin{eqnarray}}
 \newcommand\eeqn{\end{eqnarray}}

\begin{document}

\vspace*{3cm}

\centerline{\large
\bf
Diffractive
production
of Drell-Yan pairs
and
heavy
flavors}
\vspace{.5cm}

\begin{center}
 {\large
Boris~Kopeliovich}

{\sl
Max-Planck-Institut
f\"ur
Kernphysik,
Postfach
30980,
69029
Heidelberg,
Germany\\
$\&$ Joint
Institute
for
Nuclear
Research, 141980
Moscow
Region,
Dubna,
Russia}

\end{center}

\vspace{.5cm}
\begin{abstract}

Partonic interpretation of
high-energy
reactions
is known to depend on a
reference
frame. Particularly, in the rest frame of
the
target
Drell-Yan process looks 
like a freeing of the projectile $l\bar
l$
fluctuation,
rather than $q\bar q\to l\bar
l$
annihilation.
The light-cone representation for
Drell-Yan
reaction
is very similar to that in DIS and
exposes
a
substantial contamination of soft
interactions which
turns out to be dominant in
diffractive
production
of lepton pairs and in
nuclear
shadowing.
We estimate the fraction of
diffractive
events
in the total Drell-Yan cross section, which scales
in
$M^2$,
and find a substantial deviation from
factorization.
An analogous approach is developed
for
diffractive
production of
heavy
flavors.

\end{abstract}

\vspace{1.5cm}

{\large\bf Light-cone representation for Drell-Yan reaction}

\bigskip

Drell-Yan mechanism of lepton pair
production
in
hadronic
collisions at small $x_2 \ll 1$ has much
in
common
with
deep-inelastic lepton scattering at
small
Bjorken
$x_{Bj}$~\cite{hir}-\cite{bhq} (we use the standard notations \cite{dy}
$x_1-x_2=x_F$ and $x_1x_2=M^2/s$, where $x_F$ and $M$ are the Feynman
variable and the effective mass of the lepton pair). Actually, the
factorization
theorem
is a reflection of such a similarity. However,
for our purpose (diffraction, nuclear shadowing) 
it is convenient to use the
light-cone
representation for
Drell-Yan
process
\cite{hir}-\cite{bhq}.
In the rest frame of the target proton the projectile
beam
hadron
is
surrounded by a parton cloud which
contains with a small probability a
lepton pair. The freeing of the leptons by
means
of
interaction
between the projectile partons and the
target
is
equivalent
to the Drell-Yan mechanism \cite{dy} of
lepton
pair
production.

The cross section of
lepton
pair
production
can be represented in a factorized
form \cite{hir}-\cite{bhq}
\beq
M^2\frac{d\sigma_{DY}^{hN}}{dM^2dx_1}=
\int_{x_1}^1d\alpha\,
\left(\frac{x_1}{\alpha}\right)
\Phi_q^h\left(\frac{x_1}{\alpha}\right)
\,\int
d^2r_T\;
W_{ql\bar
l}\,(\alpha,r_T)\;
\sigma(q\rightarrow
ql\bar
l)\
.
\label{1}
\eeq

\noi
Here $M$ is the effective mass of the lepton pair. 
$\Phi_q^h(x)$ is the quark distribution function
in
the
projectile
hadron.
$W_{ql\bar
l}\,(\alpha,r_T)$
is
the
light-cone distribution
function
for
the
$ql\bar
l$
Fock
component of the projectile quark
in
the
mixed
$r_T-\alpha$
representation,
where $r_T$
is
the
transverse
separation between the quark and
the
center
of
gravity
of
the
$l\bar
l$-pair, and $\alpha$ is the fraction of the light-cone
momentum
of
the
projectile
quark
carried by the
$l\bar
l$
pair. The distribution
over
$r_T$
has
a form~\cite{hir,bkz}
\beq
W_{ql\bar
l}\,(\alpha,r_T)
\propto
\kappa^2K_1^2(\kappa
r_T)\
,
\label{2}
\eeq

\noi
where
\beq
\kappa^2=(1-\alpha)M^2+\alpha^2m_q^2\
.
\label{3}
\eeq

\noi
Here $m_q$ is the effective quark mass which preserves the
fluctuation from too large transverse separations compared 
to the confinement radius.
We take into account only the transversely polarized component
of the virtual photon, since the contribution of the longitudinally 
polarized one to diffraction or nuclear shadowing is a
higher twist effect.

There is a close similarity between
the
distribution
functions for a $q\bar q$ fluctuation of
a
virtual
photon~\cite{nz} and Eq.~(\ref{2}) .  This is because
the
energy
denominators corresponding to fluctuations
$\gamma^*
\to
\bar qq$ and $q \to q\bar ll$ are
very
similar.

$\sigma(q\to ql\bar l)$ in (\ref{1}) is the
cross
section
of freeing the $l\bar l$-pair fluctuation
via
interaction
with the target.  Surprisingly, it turns out to be
equal
to
the total interaction cross section with a nucleon of a colorless
$q\bar
q$
pair with a transverse separation
$\alpha
r_T$,
\beq
\sigma(q\to
ql\bar
l)=
\sigma_{\bar qq}(\alpha
r_T,x_2)\
.
\label{4}
\eeq

\noi
We included here a dependence on
$x_2$ of
the
gluon
density
in the
target. In Born approximation the
dipole
cross
section is independent of $x_2$
and
reads~\cite{zkl}
\beq
\sigma_{\bar qq}(r_T)
=
\frac{16}{3}
\int
\frac{d^2q\,\alpha_s^2}{(q^2+m_g^2)^2}\,
\left(1-
e^{i\vec
q\vec
r_T}\right)\,
\left[1-F_N(q)\right]\
.
\label{4a}
\eeq

\noi
Here $\vec q$ is the transverse momentum
of
the
exchanged
gluons,
$m_g$ is the effective gluon mass
which
takes
care of confinement, 
$F(q)=\la N|{\rm exp}[i\vec q(\vec r_1-\vec r_2)]|N\ra$ is
the
two-quark
formfactor of
the
nucleon, where $\vec r_{1,2}$ are the transverse coordinates of the
quarks coupled to the gluons.

Expression (\ref{4a})
is
infrared
stable even if $m_g=0$ due to
color
screening,
as different from the
quark-nucleon
total
cross section which
is
divergent in this case.
We should emphasize 
that
$\sigma(q\to ql\bar l)$ is not 
an interaction cross section of a
single
colored quark, but a {\it production cross section}.
The former is infra-red divergent,
but
the
latter is finite.  Indeed, a fluctuation can be
produced
on
mass shell only if the interaction amplitude
of
the
projectile Fock states with and without the
fluctuation
are
different, otherwise the coherence of the projectile
is
not
disturbed.

\begin{figure}[tbh] \includegraphics{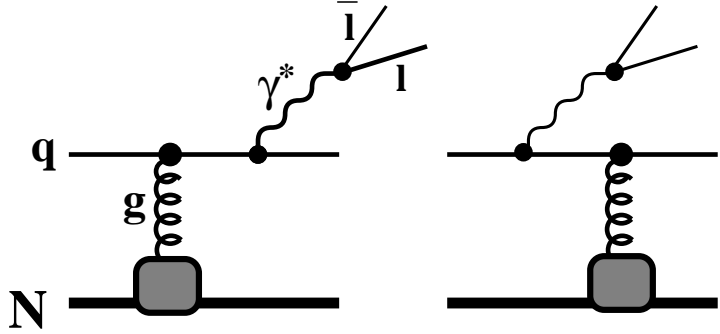}
\begin{center}
\vspace{5cm}
\parbox{13cm}
{\caption[Delta]
{\sl Feynman diagrams for lepton pair production.
The grey boxes represent the gluon distribution
function for the proton.}
\label{fig1}}
\end{center}
\end{figure}

Eq.~(\ref{1}) can also be interpreted
in
terms
of Feynman diagrams~\cite{hir,bkz} shown in Fig.~\ref{fig1}.  
The $\bar
ll$ pair production is due to
interactions
with
the target which occur prior and after the
radiation.  The impact parameters of the quark
before
and
after radiation pair are different
by
a
distance $\alpha r_T$. The corresponding
color
screening
factor turns out to be exactly the same as in
the
{\sl
total} cross section of a colorless $\bar
qq$
pair
having transverse separation
$\alpha
r_T$ \cite{hir,bkz}.

\medskip

Naively, one would expect that a typical transverse
separation
for
a
heavy $l\bar l$ fluctuation is of the order
of
$1/M^2$. This,
however,
might not be true, it depends on
$\alpha$. According
to
(\ref{2}) - (\ref{3})
the mean transverse
separation
squared
is (see
also
in~\cite{bhmt})
\beq
\langle
r_T^2\rangle
\approx
\frac{1}{(1-\alpha)\,M^2}\
.
\label{5}
\eeq

\noi
Thus, rare asymmetric $l\bar l$
fluctuations
with
$1-\alpha
\sim \mu^2/M^2$, where $\mu$ is a
hadronic
mass
scale,
interact softly. This is why the soft
contribution
to
the  
$l\bar l$ production
cross
section
is not small and scales in
$M^2$
(compare
with
DIS~\cite{kp}).
This is illustrated in Table~1 which represents
the $\bar qq$ dipole 
cross
section
$\langle\sigma\rangle$ and $\langle\sigma^2\rangle$ 
averaged over
different
fluctuations
weighted by the distribution
function
$W$ as one can read in (\ref{1}).

\vspace{.2cm}
\begin{center}
{\sl
Table
1.
Contributions of soft and hard
$l\bar
l$
fluctuations
of a
quark
\\
to the inclusive and diffractive 
cross sections of Drell-Yan
reaction.}

\vspace{.3cm}

\begin{tabular}{|c|l|l|l|l|c|}
 \hline
Fluctuation & $W_{ql\bar l}$ &
$\sigma$
&
$W_{ql\bar
l}\sigma$
&
$W_{ql\bar
l}\sigma^2$\\
\hline Hard & $\sim 1$ &
$\sim
1/M^2$
& $\sim 1/M^2$ &
$\sim
1/M^4$\\
\hline
Soft &
$\sim\mu^2/M^2$
&$\sim1/\mu^2$&
$\sim 1/M^2$ &
$\sim
1/\mu^2M^2$\\
\hline
\end{tabular}
\vspace{.7cm}
\end{center}

We classify conventionally, for the sake
of
simplicity,
all the fluctuations to be soft or hard.
The
Table
shows that soft ones have the same $M^2$
dependence
since
the rareness of their appearance is compensated by
the
large
cross section. This is a manifestation of the
aligned
jet
model of Bjorken and Kogut~\cite{bk} for
Drell-Yan
lepton
pair
production.

\bigskip

{\large\bf Nuclear shadowing}

\bigskip

This gives a hint to calculation of nuclear shadowing
for
$l\bar
l$
production. In the lowest order for multiple
scattering
expansion
it
has
a
form~\cite{kp},
\beq
\frac{\sigma(qA\to
l\bar
lX)}
{A\,\sigma(qN\to
l\bar
lX)}=
1
-
\frac
{1}{4}\,
\frac{\langle\sigma^2\rangle}
{\langle\sigma\rangle}
\langle
T\rangle\,
F^2_A(q_L)\
,
\label{6}
\eeq

\noi
The mean value of the dipole cross section is defined
as
\beq
\langle\sigma\rangle
=
\int_{0}^1d\alpha
\int
d^2r_T\;
W_{ql\bar
l}\,(\alpha,r_T)\;
\sigma_{\bar qq}(\alpha\,\cdot r_T)
\label{6a}
\eeq

The mean value $\langle\sigma^2\rangle$ is
dominated
by
soft fluctuations according to Table~1.  It has
the
same
$1/M^2$ dependence as the total
cross
section
$\langle\sigma\rangle$, therefore nuclear
shadowing
scales
in
$M^2$.

In (\ref{6}) the mean nuclear thickness
function
$\langle
T\rangle$ and the longitudinal
nuclear formfactor are defined as,
\beq
\langle
T\rangle
=\frac{1}{A}\int
d^2b\,T^2(b)\ ,
\label{6b}
\eeq

\noi
where
\beq
T(b)=\int\limits_{-\infty}^{\infty}dz\,\rho_A(b,z)\ ,
\label{6c}
\eeq
\beq
F^2_A(q_L)=\frac{1}{A\,\langle
T\rangle}\,\int
d^2b\,
\left|\int\limits_{-\infty}^{\infty}
dz\,e^{iq_Lz}\,\rho_A(b,z)\right|^2\ .
\label{7}
\eeq

\noi
The longitudinal momentum transfer $q_L$ depends
on
the
final $ql\bar l$ effective mass and approximately
equals
to
$q_L\approx 2m_Nx_2$ (see, however, \cite{krt}.  For heavy pairs $M > 4\,GeV$
in
the
experiments at SPS CERN at $p_{lab}=200\,GeV$ the value
of
$q_L$
is pretty large and the formfactor 
suppresses
nuclear shadowing.  An onset of nuclear effects
was
observed
for the first time in the experiment
E772
at
$800\,GeV$~\cite{e772} at Fermilab in a good
agreement with formula (\ref{6})~\cite{hir,bkz,krst}.

At much higher energies of RHIC and LHC one
reaches
the
regime $q_L \ll 1/R_A$  and
the
formfactor
$F^2_A(q_L) =1$.  Then the lifetime of the
$l\bar
l$
fluctuation exceeds the nuclear radius. 
In this case one can easily sum up
all
the
higher multiple scattering terms by a
simple
replacement
in~(\ref{1}),
\beq
\sigma(q\to
ql\bar
l)
\Rightarrow
2\left[1
-
\exp\left(-{1\over
2}\,
\sigma(q\to
ql\bar
l)\,T(b)\right)\right]
\label{8}
\eeq

Since nuclear shadowing for Drell-Yan reaction
is essentially eliminated 
by the nuclear formfactor in (\ref{6}) in the energy range of SPS,
this reaction can be
safely used for normalization of nuclear suppression
for $J/\Psi$ production in heavy collisions as function of centrality.
However, at higher energies, particularly at RHIC,
 the longitudinal momentum transfer
$q_L\approx 2m_NM^2/s$ in (\ref{7}) vanishes and 
Eq.~(\ref{8}) exposing a full strength of nuclear shadowing is valid .
Nuclear suppression for Drell-Yan reaction becomes quite strong 
(a factor of 0.5 or less
for collision of heavy nuclei) what makes normalization 
for nuclear suppression of $J/\Psi$ problematic.

\bigskip

{\large\bf Diffraction}

\bigskip

\begin{figure}[tbh] 
\includegraphics{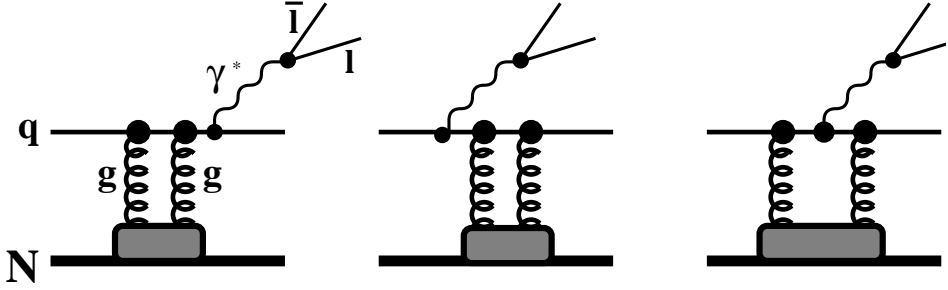}
\begin{center}
\vspace{5cm}
\parbox{13cm}
{\caption[Delta]
{\sl Feynman diagrams for diffractive production
of lepton pair. The grey boxes represent the double-gluon
distribution function for the proton.}
\label{fig2}}
\end{center}
\end{figure}

Nuclear shadowing discussed above is known
to have close
relation to diffraction~\cite{gribov}.  
Feynman diagrams for diffractive production
of a lepton pair in quark-nucleon interaction
are depicted in Fig.~\ref{fig2}.
One needs two gluons in the amplitude to insure a large
rapidity gap, which corresponds to an experimental 
definition of diffraction. 
Therefore, the diffractive cross section is related
by usual expression to the dipole cross section,
\beq
\int
dM^2\,
\left.\frac{d\sigma^{DY}_{dd}}{dM^2\,dp_T^2}
\right|_{p_T=0}
=
\frac{\langle\widetilde\sigma^2\rangle}
{16\,\pi}\ ,
\label{9}
\eeq

\noi
with only modification marked
by the {\it tilde}. Namely, one of the
exchanged gluons can
be attached to
another
projectile spectator parton (one of the
two gluons
still must couple the radiating quark \cite{gb}).
As a result, additional color screening 
suppresses the cross section $\tilde\sigma_{\bar qq}(r_T,x_2)$ as compared to
the conventional $\bar qq$-nucleon dipole cross section
$\sigma_{\bar qq}(r_T,x_2)$
\beq
\widetilde\sigma_{\bar qq}(r_T)
=
\frac{16}{3}\int
\frac{d^2q\,\alpha_s^2}{(q^2+m_g^2)^2}\,
\left(1-
e^{i\vec
q\vec
r_T}\right)\,
\left[1-F_N(q)\right]\,
\mbox{\boldmath$\left[1-F_h(q)\right]$}\ ,
\label{10}
\eeq

\noi
Where $F_h(q)$ is the double-quark formfactor of the hadron
defined in the same way as $F_N(q)$.
We use here again Born approximation, for the sake of clarity,
in analogy to (\ref{4a}). However, this expression 
differs from Eq.~(\ref{4a}) by the last factor highlighted in bold,
which takes into
account the extra color screening
of the Drell-Yan process by the spectator partons. This
is
not
a
big effect for small $r_T$ since large $q$
suppresses
the
projectile
hadron formfactor $F_h(q)$. However, diffraction,
as
was
demonstrated
above, is dominated by soft interactions, {\it
i.e.}
by
large
$r_T$ of about the size
of
the
hadron. Therefore, the last factor
in
(\ref{10})
substantially suppresses the share
of diffraction
in the total cross section compared to that
in DIS, known to be about $10\%$.
Using (\ref{10}) and
(\ref{4a}) and assuming the same $t$-slopes in
both
diffractive reactions we
get,
\beq
\left(\frac{\sigma_{dd}}
{\sigma_{tot}}\right)^{DY}
\approx
{1\over 2}\,\left(
\frac{\sigma_{dd}}
{\sigma_{tot}}\right)^{DIS}
\label{11}
\eeq

\noi
The calculation is done for a proton beam.
The difference between DIS and Drell-Yan
reaction is a direct
manifestation
of
violation of factorization
in
diffraction.
There is also another source of
deviation
from
factorization which brings an
additional
suppression
to the diffraction cross section. This is the survival
probability factor. There is an extra
condition
in diffractive
Drell-Yan reaction for the spectator partons in
the
projectile
hadron to have no
inelastic
interaction. This
can be roughly estimated in
eikonal
approximation
like it was
done
in~\cite{glm}. The result depends on energy
and ranges from few percent at the Tevatron 
collider energy up to about $20\%$ at lower energies.

\bigskip

{\large\bf Heavy flavors}

\bigskip

The light-cone dynamics of heavy flavor
production
is
very similar to that for
Drell-Yan
reaction,
except 
the pair of heavy quarks can be also radiated 
by the t-channel gluon (see Feynman diagrams in \cite{gb})
and it can participate in
the
interaction
with the
target.
The impact parameter representation for the cross section
has a form similar to (\ref{1}) \cite{hir,kst},
\beq
M^2\frac{d\sigma_{\bar QQ}^{hN}}{dM^2dx_1}=
\int_{x_1}^1d\alpha\,
\left(\frac{x_1}{\alpha}\right)
\Phi_q^h\left(\frac{x_1}{\alpha}\right)
\,\int
d^2r_T\;
W_{q\bar QQ}\,(\alpha,r_T)\;
\sigma(q\rightarrow
q\bar QQ)\ .
\label{11a}
\eeq

\noi
We use the same notations as in (\ref{1}). The distribution function
$W_{q\bar QQ}\,(\alpha,r_T)$ for a $q\bar QQ$ fluctuation
has the same form as $W_{q\bar ll}\,(\alpha,r_T)$, except replacement
$\alpha_{em}^2 \to 2\alpha_s^2/3$. The cross section of $\bar QQ$ production
in $q-N$ interaction turns out to be equal to the color dipole cross section 
for interaction of a colorless system $\bar qqg^*$ with a nucleon
(compare with (\ref{4}))
$\sigma(q\to\bar QQq)=\sigma_{\bar qqg^*}(\alpha,r_T)$
\cite{hir,kst} which has a form \cite{nz1},
\beq
\sigma_{\bar qqg^*}(\alpha,r_T)=
\frac{8}{9}\,
\left[\sigma_{\bar qq}((1-\alpha)r_T)
+
\sigma_{\bar qq}(r_T)\right]
-
{1\over 8}\,\sigma_{\bar qq}(\alpha
r_T)\ .
\label{12}
\eeq

\noi
Here the virtual gluon $g^*$ represents the color-octet $\bar QQ$ pair.
We neglect the spatial separation in the $\bar QQ$ because it is small
and the corresponding correction to the diffractive cross section considered in
\cite{npz} is suppressed by factor $1/M_{\bar QQ}^2$ (see Table~1).

To calculate diffractive production of heavy flavors one
should perform a replacement 
analogous to
(\ref{12}) for
$\widetilde\sigma(\alpha
r_T)$
in (\ref{9}) - (\ref{11}). 
The cross section of diffractive production of heavy quarks is subject
to the same strong suppression by the survival probability factor as was 
discussed above for Drell-Yan reaction.

Comparison with available data needs to take into account
the exclusive channels and the kinematical domain where the diffractive
production was observed. This will be done in a separate publication.

\bigskip

{\bf Acknowledgements}
This work was partially supported by European Network:
Hadronic Physics with Electromagnetic Probes, No FMRX CT96-0008.

\end{document}